\let\FONTS=C
\catcode`\{=1 
\catcode`\}=2 
\catcode`\#=6 
\let\x=\input
\def\y#1 {\let\input\x \let\x=\undefined \input lfonts.new}
\def\input#1 {\let\input=\y \let\y=\undefinded}
\x lplain

\input clmult01.mac
\makeatletter
\input amsfont.sty
\input cropmark.sty
\nopagecropped
\makeatother

\documentstyle[amssymb,fleqn,psfig]{clmult01}

\begin{document}
\title{Aperiodic Tilings on the Computer}
\author{Uwe Grimm and Michael Schreiber}
\institute{Institut f\"{u}r Physik, 
           Technische Universit\"{a}t Chemnitz,
           D-09107 Chemnitz, Germany}
\authorrunning{Grimm and Schreiber}
\titlerunning{Aperiodic Tilings}
\maketitle

\begin{abstract}
We briefly review the standard methods used to construct quasiperiodic
tilings, such as the projection, the inflation, and the grid method.
A number of sample {\em Mathematica}\/ programs, implementing the
different approaches for one- and two-dimensional examples, are
discussed. Apart from small examples, the corresponding programs are
not contained in the text, but are archived on the WWW.
\end{abstract}

\section{Introduction}
\label{sec:Intro}

Structure models of quasicrystals are based on tilings of space,
either on perfect quasiperiodic patterns or on random tilings.
However, besides their physical applications, many of these tilings
are also aesthetically appealing -- not by chance Penrose's original
article \cite{Penrose74} is entitled ``The R\^{o}le of Aesthetics in
Pure and Applied Mathematical Research''! Thus, even a reader who does
not plan to enter the subject of quasicrystals on a deeper level might
be interested to know how such tilings can be produced on the
computer.

The computer programs discussed here were provided for two afternoon
sessions of the summer school. Because no knowledge of particular
programming languages could be assumed, and since good graphic tools
were essential for this purpose, we decided to use an algebraic
computer package that provides all one needs. Which of the
commercially available packages one chooses is mainly a matter of
taste, and we decided in favour of {\em Mathematica}\/\footnote{{\em
Mathematica}{$^{\mbox{\circledR}}$} is a registered trademark of
Wolfram Research.} \cite{Wolfram}.  Note that the actual {\em
Mathematica}\/ programs, apart from a small example, are not
reproduced in this text -- instead, the routines are archived on the
WWW \cite{programs} and can be downloaded free of charge. At present,
this directory contains a total of six files:
\begin{itemize}
\item \verb| Read.Me |
\item \verb| ChairTiling.m |
\item \verb| FibonacciChain.m |
\item \verb| OctagonalTiling.m |
\item \verb| GridMethod.m |
\item \verb| PenrosePuzzle.m |
\end{itemize}
including some documentation, but we may choose to update or
supplement the content in the future.

During the summer school, the participants worked with {\em
Mathematica}\/ `notebooks'. The corresponding front end of {\em
Mathematica}\/ provides a rather nice programming
environment. Nevertheless, there are several reasons why we prefer to
use a standard package format \cite{Maeder} for the archived programs:
first of all, these are much smaller in size; secondly, they are less
version-dependent (the notebook format in {\em Mathematica}\/ has
changed recently, and notebooks prepared with version~3.0 of {\em
Mathematica}\/ cannot be used with earlier versions); and finally, as
these are simple ASCII files, file transfer should pose no
problems. Furthermore, if desired, our programs can of course be
included into the notebook environment by the user himself easily.

After a short survey on the different approaches employed to construct
quasiperiodic tilings, we first present a simple example -- a {\em
Mathematica}\/ program for the inflation of the so-called chair tiling
\cite{GS}. This is followed by concise descriptions of our {\em
Mathematica}\/ programs. We also briefly sketch the theory behind the
constructions, for a more detailed account we refer the reader to the
literature compiled in \cite{GrimmBaake}, and to the other chapters of
this volume.

\section{How to Construct Quasiperiodic Tilings}
\label{sec:Methods}

Whether one wants to describe the structure of real (albeit idealized)
quasicrystals, or whether one is just after a more elaborate way to
tile the bathroom floor, one faces the problem of constructing
quasiperiodic tilings. There exists a variety of mathematical
procedures \cite{Baake,Janot,Senechal}, the most popular being the
projection (or cut-and-projection) method
\cite{DuneauKatz,Elser,KKL,Kramer82,KramerNeri}. Here, one starts from
a higher-dimensional {\em periodic}\/ lattice, cuts a certain slice
out of it, and projects this slice onto a lower-dimensional (the
`physical' or `parallel') space (the remaining directions are usually
referred to as `orthogonal', `perpendicular', or `internal'
coordinates). In fact, there are several slightly different, albeit
equivalent, formulations of this method, for details see \cite{Baake}.

The simplest example -- and in fact the only one that can easily be
visualized on the computer -- is given by a projection from a periodic
two-dimensional (2D) lattice to a 1D quasiperiodic structure, the
usual toy model being the Fibonacci chain which is related to the
golden mean \mbox{$\tau=(1+\sqrt{5})/2$}. This is part of one of the
programs which is described in detail in Sect.~4.  It also contains
the description of the Fibonacci chain as a substitution sequence,
which is a common method to construct aperiodic (in general not
quasiperiodic) self-similar sequences. These serve as toy models of 1D
aperiodic order in many areas of physics \cite{BaakeGrimmJoseph}.

Inflation rules \cite{GS} provide a concept to construct self-similar,
and in particular quasiperiodic tilings in the same spirit. The rules
describe in which way each tile of a given set is decomposed into
scaled copies of the tiles, such that, after applying the rules to a
patch of a tiling, one obtains a patch consisting of the same tiles,
but with all lengths scaled by a common factor. Iterating this
procedure on some initial patch, rescaling after each step such that
the tiles stay the same, one generates larger and larger patches
approaching an infinite self-similar tiling. In Sect.~5, an example
for both the projection and the inflation approach is given for the 2D
octagonal (Ammann--Beenker) tiling
\cite{AmmannGruenbaumShephard,BaakeJoseph,DuneauMosseriOguey,KatzGratias}.

Another approach, the dualization or grid method, is due to de Bruijn
\cite{deBruijn}, see also \cite{KorepinGaehlerRhyner}: Rotating
equidistant parallel lines by angles $2\pi k/n$, $k=0,1,\ldots,n-1$,
yields an $n$-fold grid. Its dualization gives an $n$-fold tiling in
which each $p$-gon of the grid corresponds to a vertex with $p$
neighbours in the tiling. This method can very easily be translated
into a computer program, see Sect.~6, and it is also straightforward
to generalize it to higher dimensions.

Finally, quasiperiodic tilings may also be constructed via so-called
matching rules \cite{GS,Penrose74,Penrose78}. Starting with a fixed
set of proto-tiles, these give an atlas of the allowed local
configurations. If these are {\em perfect}\/ matching rules, the
quasiperiodic tiling is essentially determined; more precisely, it is
the corresponding `local indistinguishability' or `local isomorphism'
(LI) class that is fixed \cite{Baake}. On first view, this looks like
a constructive prescription that should allow to model the growth of
such structures, but this turns out to be a fallacy because it
contains an inherent non-locality, compare \cite{GrimmJoseph}.  In
practice, this means that an erroneous tile addition during
construction may only show up much later and at a completely different
place, see e.g.\ \cite{Penrose89} for a nice example. In Sect.~7, we
present a program which allows the reader to try it himself for the
rhombic Penrose tiling.

\section{A Small Example Program: Inflating the Chair Tiling}
\label{sec:Chair}

Before we address our main subject, the generation of quasiperiodic
tilings, we first present a small program to give the reader a taste
of the syntax and the structure of the programs. Clearly, we cannot
give an extensive introduction into the {\em Mathematica}\/ language,
for this the reader is referred to \cite{Maeder,Skiena,Wolfram} and to
the vast literature on {\em Mathematica}\/ (including a regularly
published journal with electronic supplements) and its applications in
science, particularly in mathematics and theoretical physics -- it is
beyond the scope of this article to list all of these. Let us only
mention a few features that we use frequently, so that even the
non-incriminated reader gets some feeling how to change the provided
{\em Mathematica}\/ programs if he wants to interfere and produce his
own tilings accordings to his own aesthetic pretensions.

The main ingredient in our programs is the manipulation of lists of
objects, as we represent a tiling in a natural way as a list of its
tiles (and the tiles, occasionally, as a list of its vertices, which
in turn are lists of their coordinates). For this, functions are used,
which are either explicitly defined (like, for instance,
\verb|f[x_]:=x^2| which defines the function $f(x)=x^2$), or are used
as so-called `pure functions'. The latter are expressions delimited by
an ampersand (\verb|&|) involving arguments \verb|#1| (or just
\verb|#|), \verb|#2|, \verb|#3|, and so on (for instance, to compute
$x^2+y^2$ one may use \verb|(#1^2+#2^2)&[x,y]| or, synonymously,
\verb|Function[#1^2+#2^2][x,y]|). Such functions can easily be applied
to the elements of a list by using the command \verb|Map|, for
example, \verb|Map[#^3&,{1,2,3,4,5}]| or, shorter,
\verb|#^3&/@{1,2,3,4,5}| results in the output \verb|{1,8,27,64,125}|.
There is also a functional operation \verb|Nest| that allows one to
apply a function iteratively, as we are going to do with the inflation
rule. For instance, \verb|Nest[f,x,3]| is equivalent to
\verb|f[f[f[x]]]|.

\begin{figure}\vspace*{-0.3pt}
\sidecaption
\mbox{\psfig{file=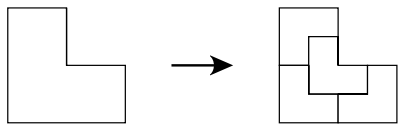,width=0.45\textwidth}}
\caption{Inflation rule of the chair tiling.}
\label{fig:chairinflation}
\end{figure}

For simplicity, we consider a tiling that consists of a single tile
only. We choose the so-called chair tiling, and generate it by
iterated application of the inflation rule given in
Fig.~\ref{fig:chairinflation}. Note that the chair tiling, albeit
aperiodic, is not quasiperiodic -- it is an example of a
limit-periodic tiling, which means that its Fourier module, though
discrete, is not finitely generated \cite{GK}. The `chair-formed'
tile, which occurs in four different orientations, can be thought of
as a combination of three congruent squares, hence the tiling belongs
to the class of so-called triomino (a term that is deduced from the
word `domino', the general case is referred to as `polyomino') tilings
\cite{Golomb}.

An implementation of the inflation rule of
Fig.~\ref{fig:chairinflation} is shown in
Fig.~\ref{fig:exampleprogram}.  \verb|ScaleFactor| gives the linear
scaling factor of the inflation rule, which is $2$ in this case. In an
inflation step, we rescale all lengths by this factor such that,
provided one starts with a patch with integer coordinates of the
vertices, the coordinates of the inflated patch are integers. A single
tile is represented in the form \verb|{or,{x0,y0}}|, where \verb|or|
denotes the angular orientation of the tile, with values $0$, $1$,
$2$, and $3$ ($\bmod 4$), and \verb|{x0,y0}| denote the coordinates of
the reference point. The function \verb|TileCoordinates| computes the
actual coordinates of the six vertices of a tile, which is used in the
graphics function \verb|PlotTiling|. All displacements are expressed
in terms of four 2D vectors, namely \verb|TwoVector[0]={1,1}| and its
$m$-times $90^{\circ}$-rotated copies \verb|TwoVector[m]|. The
inflation rule is encoded in the function \verb|TileInflation| which
produces a list of the four tiles into which a single tile is
dissected. Since tiles in all orientations are dissected in the same
way, it is sufficient to use a single definition by employing the
rotated vectors \verb|TwoVector[m]|. To perform an $n$-fold inflation
of a patch, consisting of a list of one or more tiles, the function
\verb|Inflation| is used, with $n$ as its second argument. Finally,
\verb|PlotTiling| enables one to plot the tiling, with optional
arguments that allow to give different colours to the four orientation
of the tile and to change the colour and width of the lines outlining
the tiles.

\begin{figure}
\begin{tabular}{@{}|c|@{}}
\hline
\begin{minipage}{0.94\textwidth}
\begin{small}
%
%
\begin{verbatim}


Clear[ScaleFactor,TwoVector,TileCoordinates,TileInflation,
      Inflation,PlotTiling];

ScaleFactor = 2;

TwoVector[num_Integer] := 
  TwoVector[num] = 
  Dot[MatrixPower[{{0,-1},
                   {1, 0}},num],{1,1}];

TileCoordinates[{tile_Integer,refpoint_List}] := 
  Map[(refpoint+#)&,
      {0,(#3+#4)/2,#4,#1,#2,(#2+#3)/2}]&[TwoVector[tile],
                                         TwoVector[tile+1],
                                         TwoVector[tile+2],
                                         TwoVector[tile+3]];

TileInflation[{tile_Integer,refpoint_List}] := 
  {{#1,#4},
   {#2,#4+TwoVector[#2]},
   {#1,#4+TwoVector[#1]},
   {#3,#4+TwoVector[#3]}}&[Mod[tile,4],
                           Mod[tile-1,4],
                           Mod[tile+1,4],
                           ScaleFactor*refpoint];

Inflation[tiling_List,
          num_Integer:1] /; num>=0 :=  
  Nest[Flatten[Map[TileInflation,#],1]&,tiling,num];

PlotTiling[tiling_List,
           tilecol_List:Table[GrayLevel[1],{4}],
           linecol_:GrayLevel[0],
           linewidth_:1/200] := 
  Graphics[Table[{tilecol[[i+1]],
                  Map[Polygon,#],
                  linecol,Thickness[linewidth],
                  Map[Line[Join[#,Take[#,2]]]&,#]}&[
                    Map[TileCoordinates,
                        Select[tiling,Mod[First[#],4]==i&]]],
                 {i,0,3}],
           AspectRatio->1];


\end{verbatim}
\end{small}
\end{minipage}\\
\hline
\end{tabular}\vspace*{2ex}
\caption{The {\em Mathematica}\/ program \protect\verb|ChairTiling.m|
\protect\cite{programs} for the inflation of the chair tiling. It is
also available in electronic form
\protect\cite{programs}.\label{fig:exampleprogram}}
\end{figure}

\begin{figure}[t]
\mbox{\psfig{file=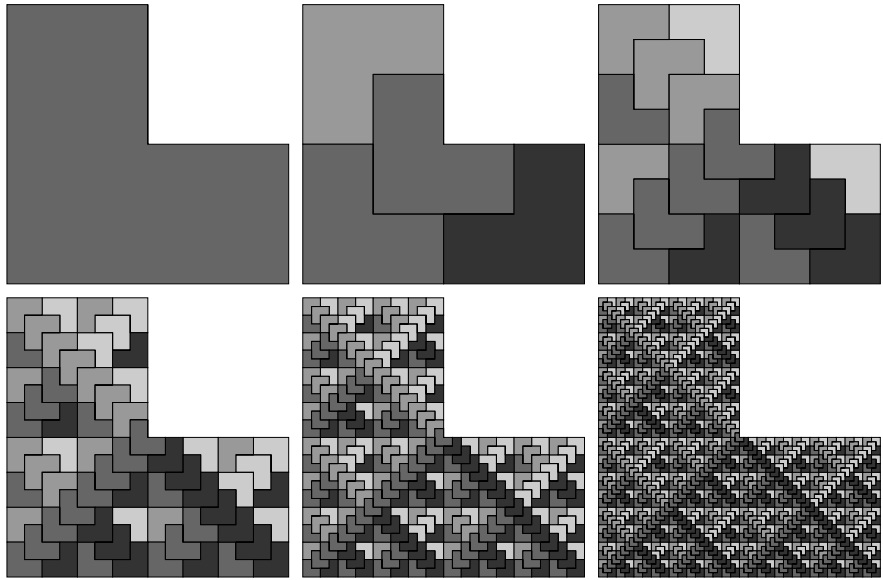,width=\textwidth}}
\caption{Inflation of the chair tiling. The four orientations of the
tile are distinguished by different shadings.}
\label{fig:chair}
\end{figure}

The result can be seen in Fig.~\ref{fig:chair}, where the first five
inflation steps of a single tile are shown, giving different shadings
to the four orientations of the tile. This figure has been produced by
the following input

{\small
\begin{verbatim}
GraphicsArray[Partition[
  Table[PlotTiling[Inflation[{{2,{0,0}}},n],
                   Map[GrayLevel,{0.8,0.6,0.4,0.2}],
                   GrayLevel[0],
                   1/500],
        {n,0,5}],3],GraphicsSpacing->0];
\end{verbatim}}

\noindent
within a \verb|Show| or \verb|Display| command to display the result
on the screen or write it to a PostScript file, respectively.  Note
that our rescaling by \verb|ScaleFactor| has effectively been reversed
by {\em Mathematica}\/, since the different subgraphs in a
\verb|GraphicsArray| arrangement are scaled to the same total
size. For completeness, we mention that Fig.~\ref{fig:chairinflation}
was produced by 

{\small
\begin{verbatim}
GraphicsArray[
  {PlotTiling[{{2,{0,0}}},Map[GrayLevel,{1,1,1,1}],
              GrayLevel[0],1/100],
   Graphics[{GrayLevel[0],Polygon[{{1/4,-1/3},{1,0},{1/4,1/3}}],
             GrayLevel[1],Disk[{1/8,0},{1/4,7/20}],
             GrayLevel[0],Thickness[1/50],Line[{{-1,0},{4/5,0}}]},
             AspectRatio->1,PlotRange->{{-2,2},{-2,2}}],
   PlotTiling[Inflation[{{2,{0,0}}}],Map[GrayLevel,{1,1,1,1}],
              GrayLevel[0],1/100]}];
\end{verbatim}}

\noindent
where the second part of the argument of \verb|GraphicsArray| produces
the arrow.

This completes our little excursion into the mysteries of {\em
Mathematica}\/.  Of course, the preceding remarks can only offer a
glimpse at the possibilities of this algebraic computer package,
although our example program of Fig.~\ref{fig:exampleprogram} -- on
purpose -- is not written in the simplest possible way, but already
contains some little tricks to show at least some of the ingredients
of the programs discussed below.

\section{Once Again: The Ubiquitous Fibonacci Chain}
\label{sec:Fibonacci}

The toy model of a 1D quasicrystal is the Fibonacci chain -- one will
hardly find any introductory text on quasicrystals where this example
is not discussed. The usual way to introduce the Fibonacci sequence
employs an inflation-like procedure, a so-called substitution rule
\begin{equation}
\varrho:\;\;\left\{\begin{array}{@{\,}lcl}
S & \rightarrow & L \\
L & \rightarrow & LS \end{array}\right.
\label{eq:substitution}
\end{equation}
on two letters $S$ and $L$, generating a semi-infinite word
$w_{\infty}$ by iterated application $w_{n+1}=\varrho(w_{n})$ on some
initial word, say $w_{0}=S$,
\begin{equation}
\begin{array}{@{}rcl@{\qquad\quad}rcl@{}}
w_{0} & = & S    & w_{4} & = & LSLLS \\
w_{1} & = & L    & w_{5} & = & LSLLSLSL \\
w_{2} & = & LS   & w_{6} & = & LSLLSLSLLSLLS \\
w_{3} & = & LSL  & w_{7} & = & LSLLSLSLLSLLSLSLLSLS \\
\end{array}\label{eq:fibonaccisequence}
\end{equation}
and so forth.  Obviously, in place of the substitution rule
(\ref{eq:substitution}), one can use concatenation of subsequent words
to generate the same sequence
\begin{equation}
w_{0}=S\, , \qquad w_{1}=L\, , \qquad w_{n+1}=w_{n}w_{n-1}\, ,
\label{eq:fibonaccirecursion}
\end{equation}
where $w_{n}w_{n-1}$ denotes the word obtained by appending $w_{n-1}$
to $w_{n}$. {}From this, it is easy to see that the length of the word
$w_{n}$ (i.e., the number of its letters) is given by the Fibonacci
number $|w_{n}|=f_{n+1}$ defined by the recursion
\begin{equation}
f_{0}=0\, , \qquad f_{1}=1\, , \qquad f_{n+1}=f_{n}+f_{n-1}\, ,
\label{eq:fibonaccinumbers}
\end{equation}
and that $w_{n}$ contains precisely $f_{n}$ letters $L$ and $f_{n-1}$
letters $S$ [for $n>0$, otherwise one has to extend the definition
(\ref{eq:fibonaccinumbers}) to negative values of $n$, which amounts
to setting $f_{-n}=(-1)^{n+1}f_{n}$ for $n>0$].  Thus, the frequency
of the predominant letter $L$ in the limit word $w_{\infty}$ is given
by
\begin{eqnarray}
\nu_{L} & = & \lim_{n\rightarrow\infty} \frac{f_{n}}{f_{n+1}}\; =\;
\lim_{n\rightarrow\infty} \frac{f_{n}}{f_{n}+f_{n-1}}\nonumber\\ 
& = & 1\, +\,\left(\lim_{n\rightarrow\infty} 
\frac{f_{n-1}}{f_{n}}\right)^{-1}\; = \; 1\, +\, \nu_L^{-1} \nonumber \\ 
\Longrightarrow \quad \nu_{L} & = & \frac{\sqrt{5}-1}{2}
\; = \; \tau\, -\, 1 \; = \; \tau^{-1}
\end{eqnarray}
where $\tau=(1+\sqrt{5})/2=\lim_{n\rightarrow\infty}(f_{n+1}/f_{n})$
is the golden ratio. As this is an irrational number, we thus proved
that the Fibonacci sequence
\begin{figure}[t]
\mbox{\psfig{file=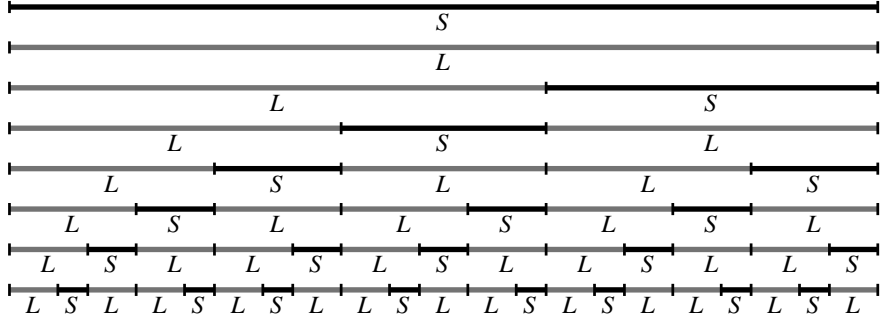,width=\textwidth}}
\caption{Inflation of the Fibonacci chain.\label{fig:fibonacci}}
\end{figure}
is aperiodic. Such statistical properties of the sequence can also be
derived from the associated substitution matrix
\begin{equation}
M_{\varrho} \; = \;\left(\begin{array}{@{\,}cc@{\,}} 
0&1\\ 1&1 \end{array}\right)\, ,
\qquad \lambda^{\pm}\; =\; \pm\tau^{\pm 1} \, , \qquad
v^{+}\; = \; \left(\begin{array}{@{}c@{}}\tau^{-2}\\ 
\tau^{-1} \end{array}\right)\, ,
\label{eq:submatrix}
\end{equation}
whose elements just count the number of letters $S$ and $L$ appearing
in $\varrho(S)$ and $\varrho(L)$, respectively. Its largest eigenvalue
$\lambda^{+}$, related to the exponential growth of the word length in
a substitution step, is just $\tau$; and the elements of the
corresponding eigenvector $v^{+}$, when properly normalized, encode
the frequencies of the two letters,
cf.~(\ref{eq:submatrix}). Obviously, substitution rules with
irrational maximum eigenvalue of the corresponding substitution matrix
generically give aperiodic (but, in general, not quasiperiodic)
chains. However, a rational or integer maximum eigenvalue does not
imply that the sequence one obtains is periodic; the Thue--Morse
sequence with substitution $a\rightarrow ab$, $b\rightarrow ba$, is a
prominent example: it is aperiodic though the two eigenvalues of the
substitution matrix (whose elements are all equal to $1$) are $2$ and
$0$.

\begin{figure}
\mbox{\psfig{file=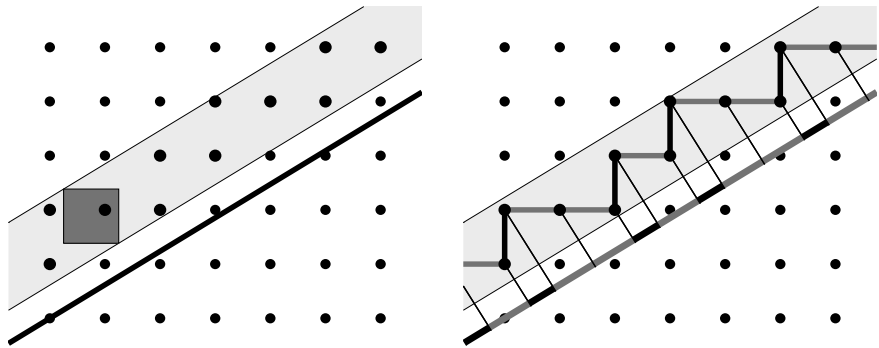,width=\textwidth}}
\caption{Projection of the Fibonacci chain from the square lattice in
the `strip projection'. The slope of the `physical space' is
$\tau^{-1}$. For details see text.\label{fig:projection1}}
\end{figure}

In the file \verb|FibonacciChain.m| \cite{programs}, a program
\verb|SubstitutionSystem| is contained that defines several functions
(\verb|SubstitutionRule|, \verb|Substitution|,
\verb|SubstitutionMatrix|, \verb|SubstitutionSequence|,
\verb|SubstitutionWord|) to generate arbitrary substitution sequences.
The substitution rule and an initial word have to be given as
parameters, the defaults (which are \verb|{"S"->{"L"},
"L"->{"L","S"}}| for the rule and \verb|{"S"}| as initial word) yield
the Fibonacci sequence as in Eqs.~(\ref{eq:substitution}) and
(\ref{eq:fibonaccisequence}).  The recursive definition
(\ref{eq:fibonaccirecursion}) is also implemented
(\verb|FibonacciRecursionSequence|,
\verb|FibonacciRecursionWord|). Here, the Fibonacci (or other
substitution) words may be represented in two ways; for instance,
$w_{4}=LSLLS$ may either be represented as a list of single letters
\verb|{"L","S","L","L","S"}| (e.g.,
\verb|FibonacciRecursionSequence[4]|) or as a string \verb|{"LSLLS"}|
(e.g., \verb|FibonacciRecursionWord[4]|). There are a number of
functions to analyze the statistical properties of the sequence which
are written such that they can simultaneously deal with both
representations of the words. For example, \verb|SubWordCount[w1,w2]|
counts the number of occurrences of the word \verb|w2| in the word
\verb|w1|. Furthermore, the Fibonacci numbers are defined
(\verb|FibonacciNumber[n]| gives $f_n$), and a small program
\verb|PlotLinearChain| for a geometric representation of $n$-letter
sequences as linear arrangements of $n$ intervals is included. This we
used to prepare Fig.~\ref{fig:fibonacci} by

{\small
\begin{verbatim}
PlotLinearChain[Table[SubstitutionSequence[i],{i,0,7}],
                True,
                {N[GoldenRatio],1},
                {GrayLevel[0.45],GrayLevel[0]},
                {GrayLevel[0],GrayLevel[0]}]
\end{verbatim}}

\noindent where the length ratio of the two intervals is the golden
mean $\tau$, with $L$ corresponding to the `long` and $S$ to the
`short' interval, respectively. In this way, the substitution rule may
be interpreted as an inflation procedure for the geometric object;
and, upon suitable rescaling of lengths (i.e., by a factor of $\tau$),
the semi-infinite line representing the limit word $w_{\infty}$ is a
fixed point under this procedure.

\begin{figure}[t]
\psfig{file=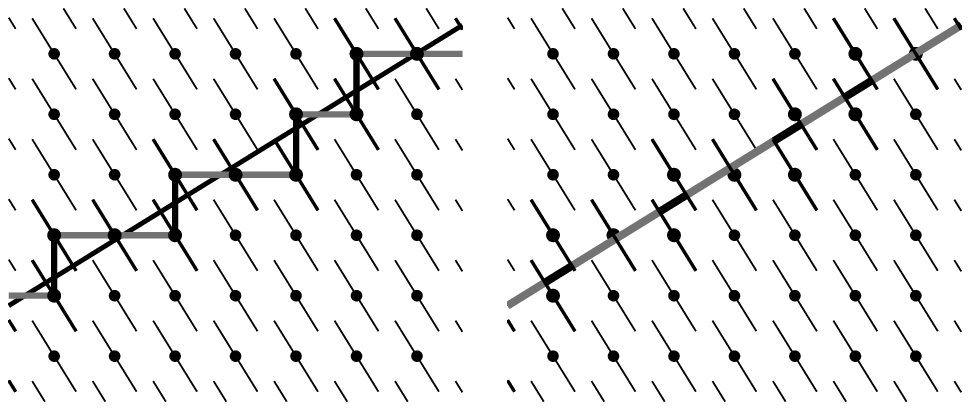,width=\textwidth}\vspace*{2ex}
\psfig{file=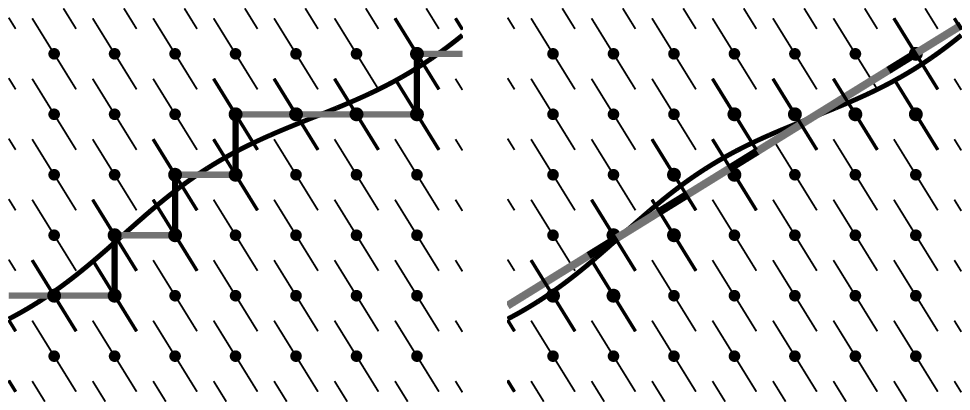,width=\textwidth}
\caption{Projection of the Fibonacci chain from the square lattice in
the `atomic hypersurface' scheme. Below, small `phasonic' fluctuations
in the cut space lead to `phason jumps', creating subwords ($LSLSLS$
and $LLL$) that are not present in a perfect Fibonacci
sequence.\label{fig:projection2}}
\end{figure}

\begin{figure}
\mbox{\psfig{file=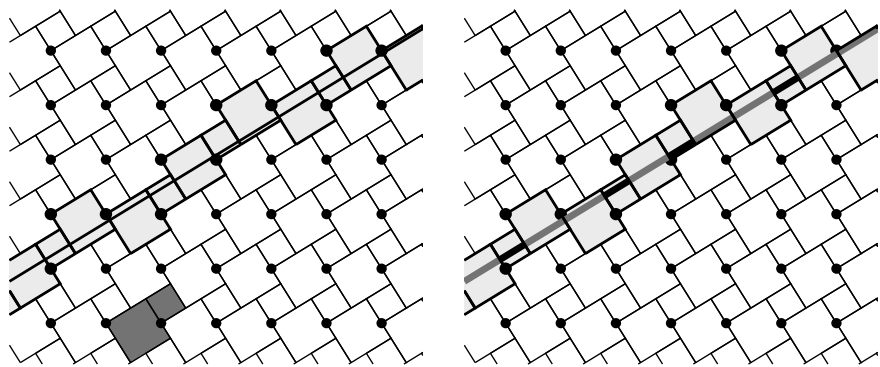,width=\textwidth}}
\caption{Projection of the Fibonacci chain from the square lattice in
the `Klotz construction' or `dualization'
scheme.\label{fig:projection3}}
\end{figure}

This `geometric Fibonacci chain', however, can also be obtained by a
suitable projection from the 2D square lattice. As mentioned
previously, there are several variants of projection schemes, which
are essentially equivalent, see Figs.~\ref{fig:projection1},
\ref{fig:projection2}, and \ref{fig:projection3}. The file
\verb|FibonacciChain.m| \cite{programs} contains the three functions
\verb|StripProjection|, \verb|AtomicSurfaceProjection|, and
\verb|KlotzConstruction|, which show, in a sequence of pictures, three
variations of the projection to a line. Among others, the slope can be
given as a parameter, the default value $1/\tau$ again yields the
Fibonacci chain.  As commonly, the `window' or `acceptance domain',
determining which part of the square lattice is projected, is chosen
as the projection of a unit cell (e.g.\ the dark grey square in the
left part of Fig.~\ref{fig:projection1}) of the square lattice onto
`perpendicular space', i.e., into the direction orthogonal to the cut
line which corresponds to the `physical space' and is indicated by the
thick line in the left part of Fig.~\ref{fig:projection1}. In the
`strip projection' shown in Fig.~\ref{fig:projection1}, this
determines the width of the strip, its vertical position can be chosen
as a parameter. The actual sequence obtained by projection depends on
this parameter, but one always stays within the same LI class
\cite{Baake}, except for some `singular' cases which are obtained if
one lattice point falls precisely on the boundary of the strip.  In
the method of the `atomic (hyper)surfaces' \cite{Bak,JannerJanssen}
shown in Fig.~\ref{fig:projection2}, the cut line is dissected by
'atomic surfaces' attached to the vertices of the square lattice.  The
function \verb|AtomicSurfaceProjection| also allows to introduce
fluctuations of the cut space (mimicking `phason' degrees of freedom,
see \cite{Trebin}), an example is shown in the lower part of
Fig.~\ref{fig:projection2}. Finally, the `Klotz construction' or
`dualization' scheme
\cite{Kramer,KramerSchlottmann,OgueyDuneauKatz,Schlottmann} shown in
Fig.~\ref{fig:projection3} employs a different unit cell of the square
lattice (e.g.\ consisting of the two dark grey squares in the left
part of Fig.~\ref{fig:projection3}), chosen such that all its
boundaries are parallel or perpendicular to the `physical space'. In
addition to the pictures shown in Fig.~\ref{fig:projection3}, the
function \verb|KlotzConstruction| also gives a detailed description of
the construction of this unit cell.

We now turn our attention to arguably more interesting examples of 2D
quasiperiodic tilings, both from an aesthetic and a physical point of
view. Nevertheless, the methods we are going to use are very similar
to the 1D case considered above, particularly the projection scheme is
a straightforward generalization of the method shown in
Figs.~\ref{fig:projection1}, \ref{fig:projection2}, and
\ref{fig:projection3}.

\section{The Octagonal Tiling: Projection and Inflation}
\label{sec:Oct}

\begin{figure}[h]
\sidecaption \mbox{\psfig{file=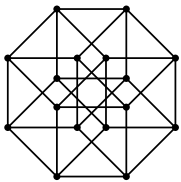,width=0.2\textwidth}}
\caption{Projection of the 4D hypercube.\label{fig:cube}}
\end{figure}

We now consider the canonical octagonal (also known as
Ammann--Beenker) tiling
\cite{AmmannGruenbaumShephard,BaakeJoseph,DuneauMosseriOguey,KatzGratias}.
The corresponding programs are contained in the file
\verb|OctagonalTiling.m| \cite{programs}. The tiling can be obtained
by projection from the 4D hypercubic lattice, choosing the orientation
of the `physical space' as one of the two unique invariant subspaces
with respect to the action of the dihedral group $D_8$ (the symmetry
group of the regular octagon), which is a subgroup of the point group
of the 4D hypercubic lattice. The acceptance domain is the orthogonal
projection of the hypercube, it is a regular octagon as shown in
Fig.~\ref{fig:cube}, which was produced with the function
\verb|PlotCubeOrthogonalProjection|.

\begin{figure}[t]
\sidecaption
\mbox{\psfig{file=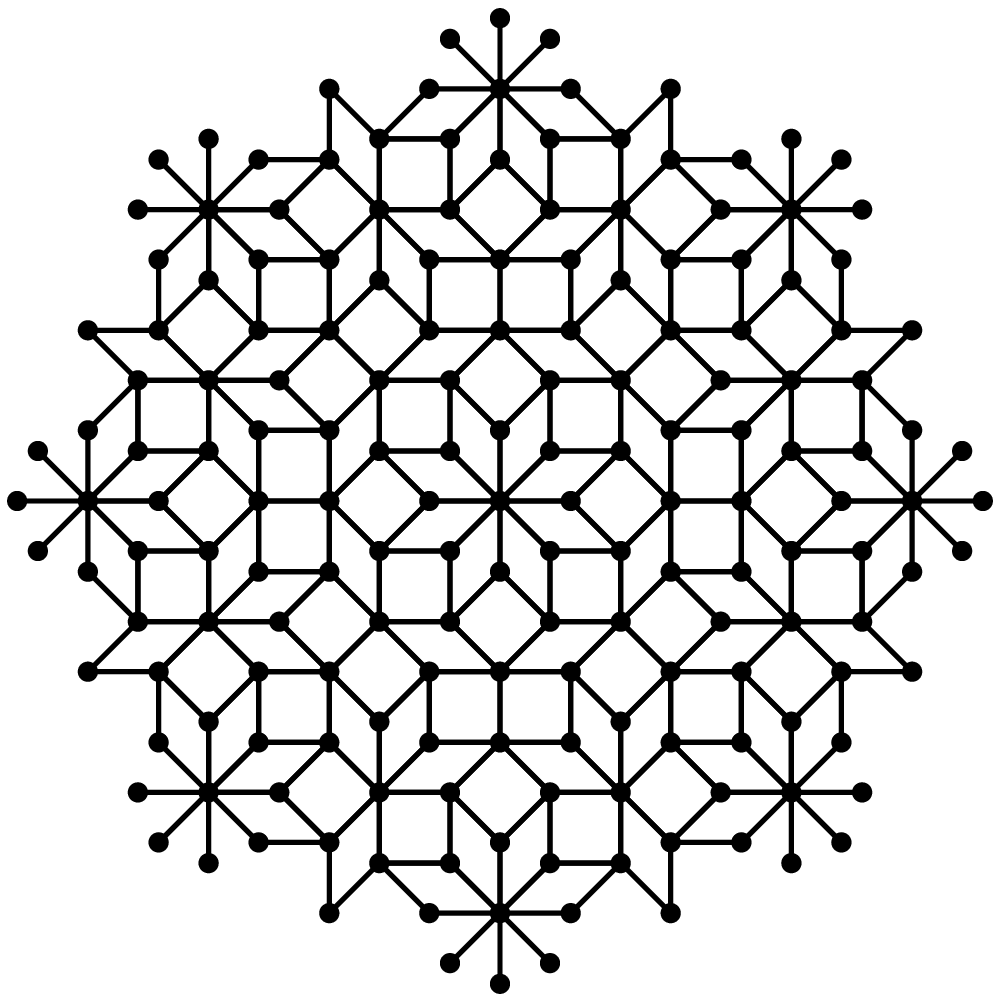,height=0.3\textheight}\hspace*{0.05\textwidth}%
\psfig{file=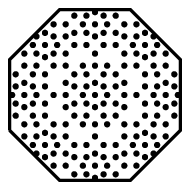,height=0.3\textheight}}
\caption{Parallel (vertices and edges, left side) and perpendicular
(vertices only, right side) projection of part of the hypercubic
lattice, drawn at the same scale. The octagon is the projection of the
4D hypercube, cf.~Fig.~\protect\ref{fig:cube}.\label{fig:projection}}
\vspace*{-1ex}
\end{figure}

The projection of a small portion of the hypercubic lattice, together
with the orthogonal projection of the vertices, is shown in
Fig.~\ref{fig:projection}. The patch comprises 177 vertices and 536
edges; it was obtained by considering all points that can be reached
by at most eight steps along the hypercubic edges from the origin
which corresponds to the center of this eightfold symmetric patch.
Here, we used the commands \verb|PlotParallelProjection| and
\verb|PlotOrthogonalProjection| defined in \verb|OctagonalTiling.m|
\cite{programs}.

\begin{figure}[b]
\vspace*{-1ex}
\mbox{\psfig{file=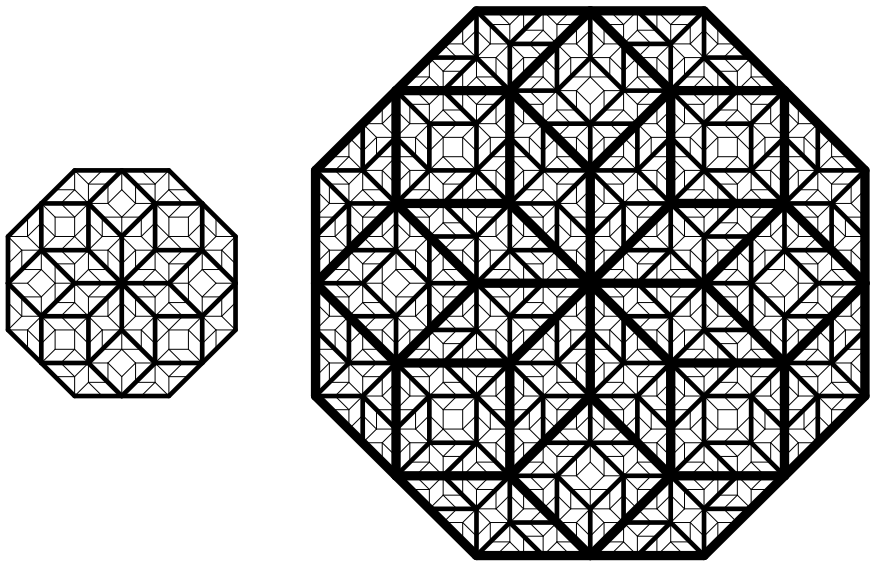,width=\textwidth}}
\caption{Two (left) and three (right) inflation steps of
the octagonal tiling.\label{fig:octagonalinflation}}
\end{figure}

The octagonal tiling can also be obtained by inflation. The inflation
rule, i.e., the description of how to cut tiles into smaller pieces,
can be read off from Fig.~\ref{fig:octagonalinflation}, which shows
three subsequent inflation steps.  Apparently, the inflation rule is
rather complicated: first of all, the parts obtained by inflating a
tile cover sometimes only half a tile at the smaller scale, and
secondly, the square tile is dissected in an asymmetric way.  The best
way to represent the inflation rule is therefore in terms of two
triangular tiles obtained by cutting the rhomb along the short
diagonal, and the square such that the two halves, when oriented
properly, are dissected in the same way. This inflation procedure has
been implemented in \verb|OctagonalInflation|, and only in the end,
when using \verb|PlotOctagonalTiling| to display the tiling, one may
choose to recombine the triangles into squares and rhombs as it has
been done in Fig.~\ref{fig:octagonalinflation}.

In the programs, the tilings are always treated in terms of integer
coordinates, only for the purpose of the graphical presentation it is
necessary to translate these into real 2D coordinates. In the
projection approach, it is natural to use the coordinates of the 4D
hypercubic lattice point, whereas in the inflation approach, we
instead performed all calculations in the module
${\Bbb{Z}}+\sqrt{2}{\Bbb{Z}}$.

\section{De Bruijn's Ingenuity: The Dualization Method}
\label{sec:Dual}

\begin{figure}[b]
\mbox{\psfig{file=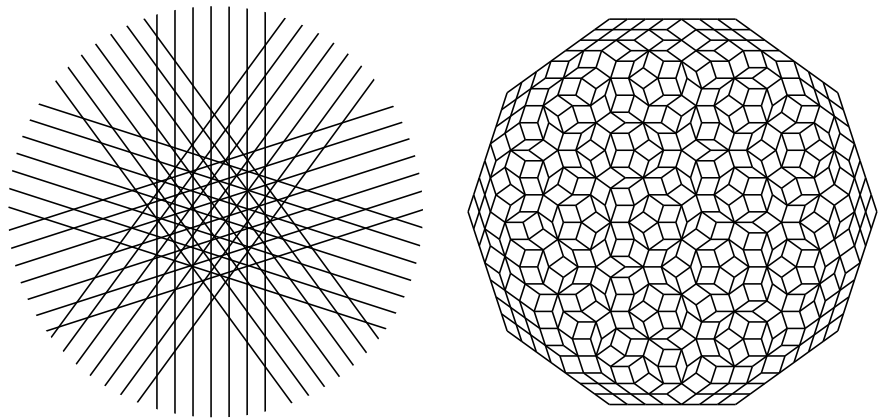,width=\textwidth}}
\caption{A pentagrid (left) and the corresponding part of a ten-fold
tiling (right, disregarding the boundary region), which belongs to the
LI class of the Penrose tiling ($\Gamma=0$).\label{fig:grid5}}
\end{figure}

Another method that is particularly easy to program is de Bruijn's
famous grid (or dualization) method
\cite{deBruijn,KorepinGaehlerRhyner}.  Here, one starts from an
$n$-fold grid that is created rotating a set of equidistant parallel
lines by angles $2\pi k/n$, $k=0,1,\ldots,n-1$, and shifting them by
certain amounts. The sum of these shifts, $\Gamma$, plays an important
role -- in general, the tilings constructed with different values of
$\Gamma$ (modulo 1) belong to different LI classes. A grid is called
`regular' if there are no intersection points where more than two
lines meet. {}From this, an $n$-fold tiling is obtained by
dualization, so that each $p$-gon of the grid yields a vertex with $p$
neighbours in the tiling. Its coordinates can be calculated
easily. For a regular grid, each intersection point of the grid thus
corresponds to a rhombic tile.  Note that for odd values of $n$ one
could as well have considered the case $2n$ since a rotation by an
angle of $\pi$ is irrelevant.

\begin{figure}[t]
\mbox{\psfig{file=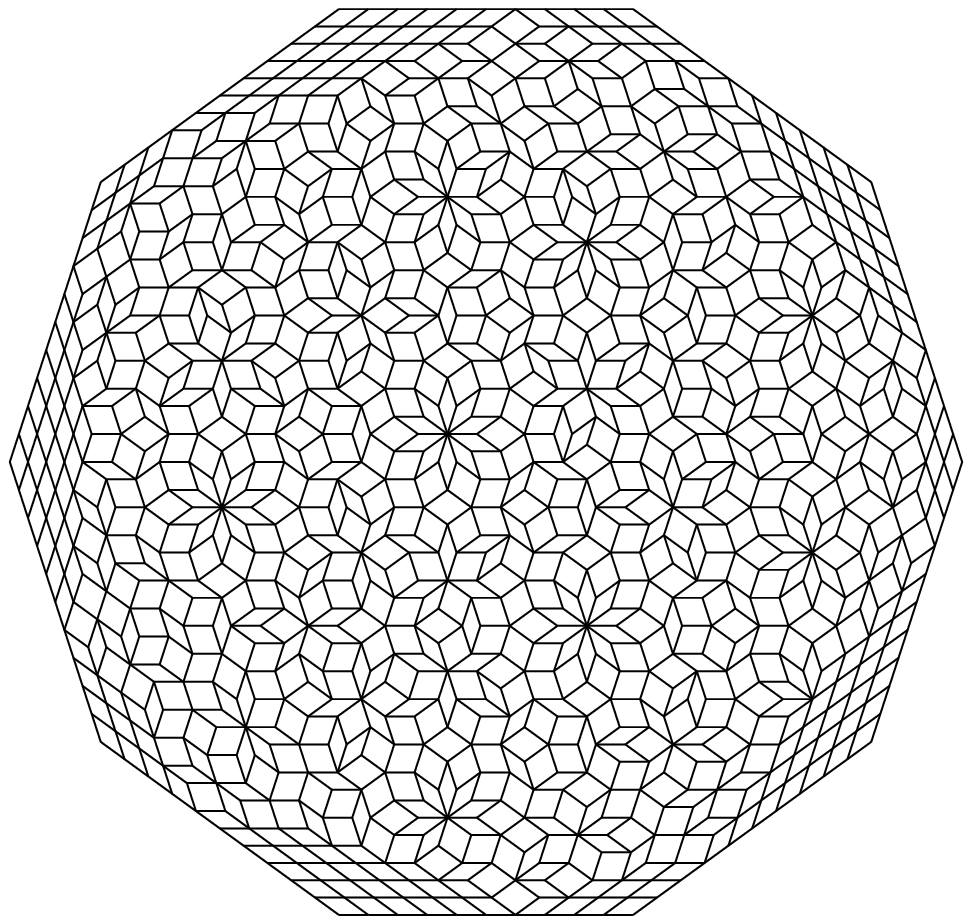,width=0.48\textwidth}\hspace{0.04\textwidth}%
\psfig{file=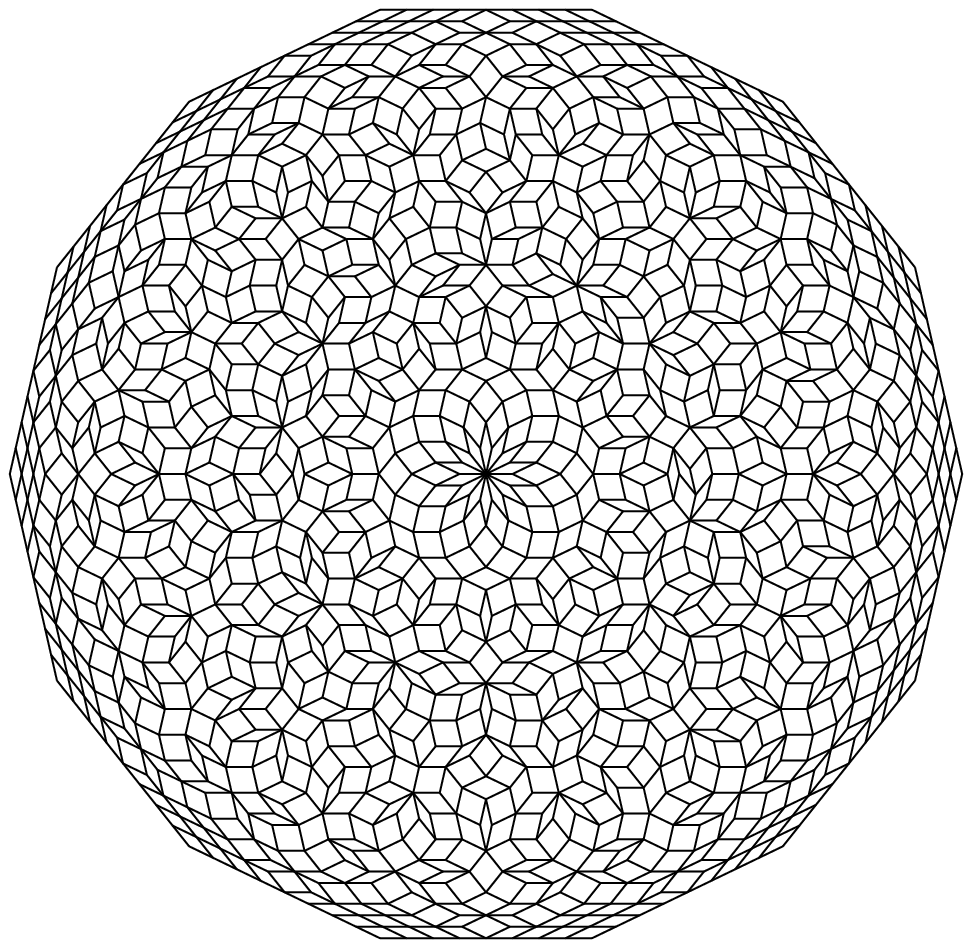,width=0.48\textwidth}}
\caption{Two tilings obtained by the grid method: a ten-fold tiling
(left) in the LI class of the `anti-Penrose' tiling ($\Gamma=1/2$) and
an example for $n=7$ (right), again disregarding the boundary
region which is due to the finite grid.\label{fig:grid57}}
\end{figure}

In Fig.~\ref{fig:grid5}, we present a grid and the corresponding
rhombic tiling for the Penrose case, i.e., for $n=5$ where the shifts
(grid parameters) satisfy the condition $\Gamma=0$. The portion of a
regular lattice on the boundary of the patch is due to the finiteness
of the grid (because we used only seven grid lines for each
direction). The picture has been produced with the commands
\verb|PlotGrid| and \verb|PlotDualTiling| defined in
\verb|GridMethod.m| \cite{programs}. Note that in the dualization,
performed in \verb|DualizeGrid|, the regularity of the grid is not
verified, although it is assumed in the construction, hence results
are unpredictable in case the grid is not regular.
Fig.~\ref{fig:grid57} shows, on the left, another example for $n=5$,
but now with $\Gamma=1/2$ which is `maximally remote' from the Penrose
case $\Gamma=0$ wherefore it is occasionally referred to as an
`anti-Penrose tiling'. Though it is built from the same two rhombic
tiles as the Penrose tiling, it obviously belongs to a different LI
class because it contains vertex configurations forbidden in a perfect
Penrose tiling, as for instance the star of ten `thin' rhombs. On the
right of Fig.~\ref{fig:grid57}, an example for $n=7$ is shown. We note
that the grid method is equivalent to the projection scheme, see
e.g.~\cite{Schlottmann} for details.

\section{Giving it a Trial: A Penrose-Tiling Puzzle}
\label{sec:Puzzle}

\begin{figure}[t]
\sidecaption \mbox{\psfig{file=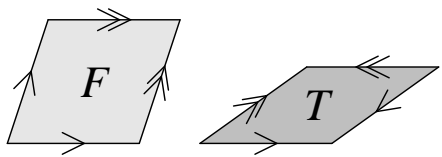,width=0.5\textwidth}}
\caption{The two Penrose rhombs (`fat' and `thin') with arrow
decorations encoding the matching rules.\label{fig:rhombs}}
\end{figure}

Finally, we like to invite the reader to take the part of the pitiable
craftsman who, unsuspectingly, accepted the honourable commission to
tile a hypocrite quasicrystallographer's bathroom floor with a portion
of a rhombic Penrose tiling.  Prudently, the quasicrystallographer
provided two sets of rhombic tiles with edges that only fit in certain
ways, thus encoding the matching rules of the Penrose tiling as in
Fig.~\ref{fig:rhombs}.

To understand the desperate situation of the tiler, the reader should
use the program \verb|PenrosePuzzle| contained in the file
\verb|PenrosePuzzle.m| \cite{programs}. One may give an arbitrary
initial patch as input, default is a single `fat' rhomb. Once started,
the program asks whether to add or to remove tiles.  An addition is
specified by the letter \verb|A| followed by the type of the new tile
(\verb|F| for the `fat' and \verb|T| for the `thin' rhomb,
respectively) and by the number of the surface edge at which the new
tile should be attached.  
\begin{figure}[b]
\mbox{\psfig{file=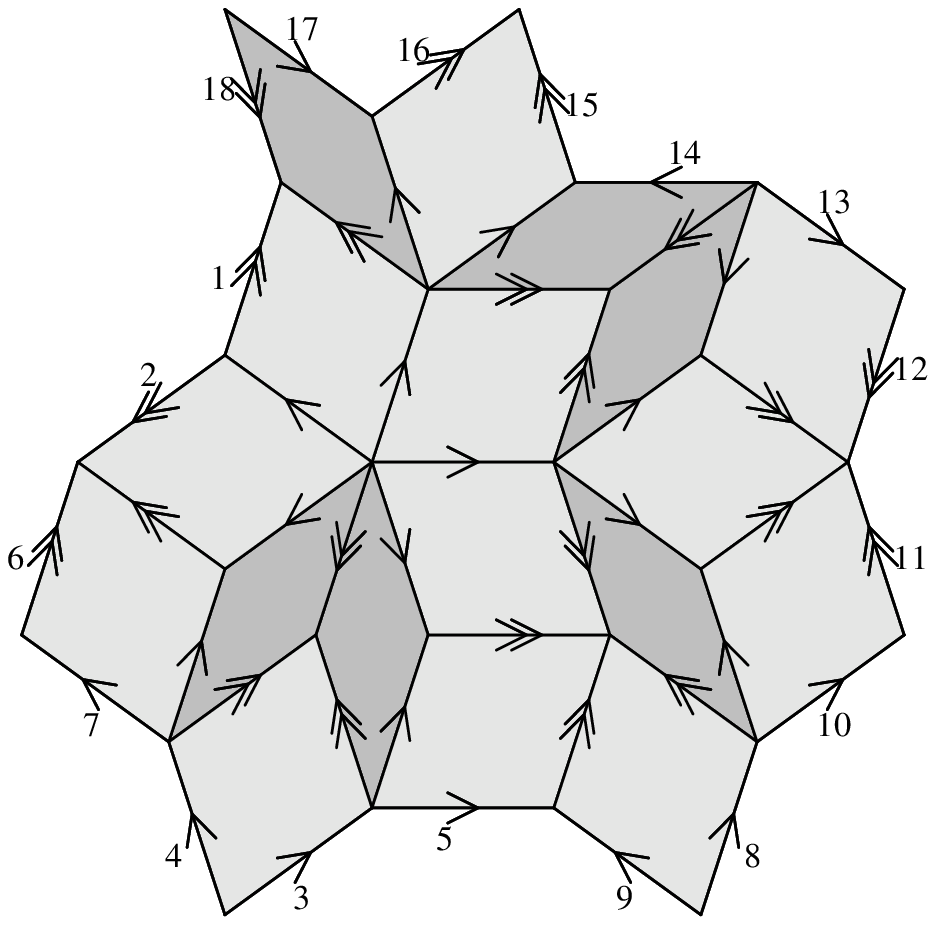,width=0.5\textwidth}%
\psfig{file=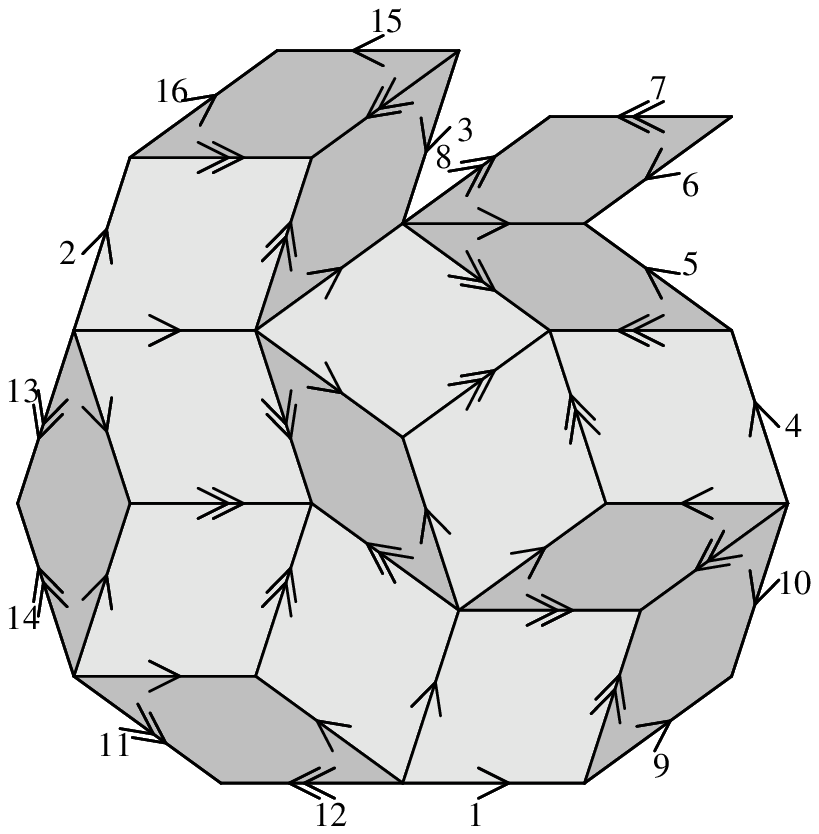,width=0.5\textwidth}}
\caption{Two patches consisting of Penrose rhombs, grown with
\protect\verb|PenrosePuzzle|. In the right example, the two crevices
with surfaces edges $(5,6)$ and and $(3,8)$ cannot be filled with the
tiles of Fig.~\protect\ref{fig:rhombs} -- any attempted addition of a
tile will produce an error due to a mismatch of the arrows or an
overlap of two tiles. Note that it does not suffice to remove the tile
with surface edges $6$, $7$ and $8$ -- one also has to remove the thin
rhomb with surface edge $5$ in order to obtain a patch that occurs in
a perfect Penrose tiling.\label{fig:patches}}
\end{figure}
This leaves only one possibility to attach the tile without creating
an overlap with the tile that it is attached to; and for each addition
the program checks whether it leads to an arrow mismatch in the
resulting tiling, or whether an overlap between the new tile and a
tile of the previous patch occurs. If that happens, an error message
is produced and one may choose whether to continue with the
addition. A tile removal is simply specified by the letter \verb|R|
together with the number of one of the surface edges of the tile to
remove (thus only tiles that are on the surface of the patch can be
removed). The edges are numbered consecutively, the patch with the
edge labels is shown after each successful addition or removal of a
tile, see Fig.~\ref{fig:patches}.

Of course, matching rules just state that one constructed a perfect
Penrose tiling when one succeeded to tile the entire plane without
violating the rules. However, since both tiles and bathrooms are
usually of finite size, the tiler might in the end turn up with a
finite `legal' patch (i.e., the matching rules are not violated) that
nevertheless is not `correct', i.e., it does not occur in a perfect
Penrose tiling. In other words, it is impossible to continue the
addition of further tiles without violating the matching rules at some
point, see the right part of Fig.~\ref{fig:patches} for a simple
example and \cite{GrimmJoseph,Penrose89} for more details.

\section{Concluding Remarks}
\label{sec:Conc}

Clearly, the programs introduced here form by no means a complete
package that allows one to create all those tilings that
quasicrystallographers play around with -- but this was not what we
had intended. Instead, it was our purpose to show with a few
comparatively simple examples how quasiperiodic tilings can be
produced, and thereby give insight into the different approaches that
one may choose. In particular, we restricted ourselves to one- and 2D
examples, not because it would be substantially more difficult to
generate higher-dimensional tilings, but mainly because it requires
much more effort to visualize the result on a 2D screen. For the 3D
case, commercially available balls-and-sticks kits, e.g.\ \cite{Zome},
which allow to build icosahedral structures, may still prove more
useful than computer-generated pictures -- however, there is also a
{\em Mathematica}\/ implementation of the balls-and-sticks tools
\cite{ElserBS}.

Nevertheless, our sample programs \cite{programs} may serve as a
starting point for the layman who aims to generate some beautiful
pictures of tilings, his interest be of mathematical, physical, or
purely aesthetical nature.  Whether experts may still profit from our
small example programs, they have to judge by themselves -- we did not
spend too much effort to find the most elegant or most concise (but
probably also quite unreadable) form of the program code, neither did
we try to optimize our programs with respect to CPU time or memory
consumption (which sooner or later -- more likely sooner -- becomes
the limiting factor in using {\em Mathematica}\/). Still, it should be
rather easy to modify our programs in order to treat most of the
quasiperiodic tilings one may encounter in the literature.

Finally, we like to note that computer programs generating
quasiperiodic tilings can also be found in other places. For example,
some {\em Mathematica}\/ programs are contained in the book
\cite{Senechal}, and there even exists an interactive WWW site
\cite{QuasiTiler} devoted to quasiperiodic tilings.


\begin{thebibliography}{References}{}
%
%
\bibitem{AmmannGruenbaumShephard}%
Ammann, R., Gr\"{u}nbaum, B., Shephard, G.C.\ (1992):
Discrete Comput.\ Geom.\ {\bf 8}, 1
%
%
\bibitem{Baake}%
Baake, M.\ (in this volume)
%
%
\bibitem{BaakeGrimmJoseph}%
Baake, M., Grimm, U., Joseph, D.\ (1993):
Int.\ J.\ Mod.\ Phys.\ B {\bf 7}, 1527 
%
%
\bibitem{BaakeJoseph}%
Baake, M., Joseph, D.\ (1990):
Phys.\ Rev.\ B {\bf 42}, 8091
%
%
\bibitem{Bak}%
Bak, P.\ (1986):
Scripta Met.\ {\bf 20}, 1199
%
%
\bibitem{deBruijn}%
de Bruijn, N.G.\ (1981):
Proc.\ Kon.\ Ned.\ Akad.\ Wet.\ A (Indagationes Mathematicae)
{\bf 84}, 39 and 53
%
%
\bibitem{DuneauKatz}%
Duneau, M., Katz, A.\ (1985):
Phys.\ Rev.\ Lett.\ {\bf 54}, 2688
%
%
\bibitem{DuneauMosseriOguey}%
Duneau, M., Mosseri, R., Oguey, C.\ (1989):
J.\ Phys.\ A {\bf 22}, 4549
%
%
\bibitem{QuasiTiler}%
Durand, E.\ (1994): {\em QuasiTiler 3.0}\/
(WWW front end by P.\ Burchard, D.\ Meyer, and E.\ Durand,
Geometry Center, University of Minnesota)
http://www.geom.umn.edu/apps/quasitiler
%
%
\bibitem{Elser}%
Elser, V.\ (1986): Acta Cryst.\ A {\bf 42}, 36
%
%
\bibitem{ElserBS}%
Elser, V.\ (1998): {\em IQtools}\/ 
(for detailed information, please contact the author by email: 
ve10@cornell.edu)
%
%
\bibitem{GK}%
G\"ahler, F., Klitzing, R.\ (1997):
{\em The Mathematics of Long-Range Aperiodic Order} 
(NATO ASI C 489), ed.\ by R.V.~Moody, Kluwer, Dordrecht, p.~141
%
%
\bibitem{Golomb}%
Golomb, S.\ (1994):
{\em Polyominoes: Puzzles, Patterns, Problems and Packings}.
(2nd ed.) Princeton University Press, Princeton
%
%
\bibitem{GrimmBaake}%
Grimm, U., Baake, M.\ (in this volume)
%
%
\bibitem{GrimmJoseph}%
Grimm, U., Joseph, D.\ (in this volume)
%
%
\bibitem{programs}%
Grimm, U., Schreiber, M.\ (1998): 
{\em Constructing Aperiodic Tilings with Mathematica: 
Some Example Programs},\newline
\mbox{http://www.tu-chemnitz.de/ftp-home/pub/Local/physik/AperiodicTilings/}
%
%
\bibitem{GS}%
Gr\"{u}nbaum, B., Shephard, G.C.\ (1987):
{\em Tilings and Patterns}, 
W.H.\ Freeman, New York
%
%
\bibitem{Janot}%
Janot, C.\ (1994):
{\em Quasicrystals: A Primer}\/ (2nd ed.), 
Clarendon Press, Oxford
%
%
\bibitem{JannerJanssen}%
Janner, A., Janssen, T.\ (1986):
Phys.\ Rev.\ B {\bf 15}, 643
%
%
\bibitem{KKL}%
Kalugin, P.A., Kitayev, A.Y., Levitov, L.S.\ (1985):
JETP Lett.\ {\bf 41}, 145; 
J.\ Phys.\ Lett.\ (France) {\bf 46}, L601
%
%
\bibitem{KatzGratias}%
Katz, A., Gratias, D.\ (1994):
{\em Lectures on Quasicrystals},
ed.\ by F.~Hippert and D.~Gratias,
Les Editions de Physique, Les Ulis, p.~187
%
%
\bibitem{KorepinGaehlerRhyner}%
Korepin, V.E., G\"{a}hler, F., Rhyner, J.\ (1988):
Acta Cryst. A {\bf 44}, 667
%
%
\bibitem{Kramer82}%
Kramer, P.\ (1982):
Acta Cryst.\ A {\bf 38}, 257
%
%
\bibitem{Kramer}%
Kramer, P.\ (1987):
Mod.\ Phys.\ Lett.\ B {\bf 1}, 7
%
%
\bibitem{KramerNeri}%
Kramer, P., Neri, R.\ (1984):
Acta Cryst.\ A {\bf 40}, 580; 
{\bf 41}, 691 (1985) [Erratum]
%
%
\bibitem{KramerSchlottmann}%
Kramer, P., Schlottmann, M.\ (1989):
J.\ Phys.\ A {\bf 22}, L1097
%
%
\bibitem{Maeder}%
Maeder, R.\ (1990):
{\em Programming in Mathematica}\/,
Addison--Wesley, Redwood City (California)
%
%
\bibitem{OgueyDuneauKatz}%
Oguey, C., Duneau, M, Katz, A.\ (1988):
Commun.\ Math.\ Phys.\ {\bf 118}, 99
%
%
\bibitem{Penrose74}%
Penrose, R.\ (1974):
Bull.\ Inst.\ Math.\ Applications {\bf 10}, 266
%
%
\bibitem{Penrose78}%
Penrose, R.\ (1978):
Eureka {\bf 39}, 16; reprinted in: 
Math.\ Intell.\ {\bf 2}, 32 (1979)
%
%
\bibitem{Penrose89}%
Penrose R.\ (1989):
{\em Introduction to the Mathematics of Quasicrystals}\/
(Aperiodicity and Order, vol.~2),
ed.\ by M.V.\ Jari\'{c},
Academic Press, San Diego, p.~53
%
%
\bibitem{Senechal}%
Senechal, M.\ (1995):
{\em Quasicrystals and Geometry},
Cambridge University Press, Cambridge
%
%
\bibitem{Schlottmann}%
Schlottmann, M.\ (1993): 
Int.\ J.\ Mod.\ Phys.\ B {\bf 7}, 1351; and
{\em Geometrische Eigenschaften quasiperiodischer Strukturen},
PhD Thesis, Universit\"at T\"ubingen
%
%
\bibitem{Skiena}%
Skiena, S.\ (1990):
{\em Implementing Discrete Mathematics: 
Combinatorics and Graph Theory with Mathematica},
Addison--Wesley, Redwood City (California)
%
%
\bibitem{Trebin}%
Trebin H.-R.\ (in this volume)
%
%
\bibitem{Wolfram}%
Wolfram, S.\ (1991): 
{\em Mathematica: A System for Doing Mathematics by Computer}\/ 
(2nd edition),
Addison--Wesley, Reading (Massachusetts)
%
%
\bibitem{Zome}%
Zometool Marketing, 
1526 South Pearl Street, Denver, CO 80210, USA;
http://www.zometool.com
%
%
\end{thebibliography}
\end{document}